\documentclass[sigconf]{acmart}

\AtBeginDocument{%
  }

\usepackage{listings}
\usepackage{amsmath,amsfonts}
\usepackage{algorithmic}
\usepackage{graphicx}
\usepackage{textcomp}
\usepackage{xcolor}
\usepackage{booktabs}
\usepackage{hyperref}
\definecolor{codegreen}{rgb}{0,0.6,0}
\definecolor{codered}{rgb}{0.6,0,0}
\definecolor{codegray}{rgb}{0.5,0.5,0.5}
\definecolor{codepurple}{rgb}{0.58,0,0.82}
\definecolor{backcolour}{rgb}{255, 255, 255}
\definecolor{framecolor}{rgb}{0.7,0.7,0.7} % light grayish color for the frame
\usepackage{tablefootnote}
\usepackage{glossaries}
\makeglossaries
\usepackage{marginnote}
\usepackage[normalem]{ulem}
\usepackage{balance}

\usepackage{xr}
\usepackage{multirow}
\usepackage{array}
\usepackage{soul}  % for \hl{}
\usepackage{tcolorbox}

\lstdefinestyle{mystyle}{
  backgroundcolor=\color{backcolour}, 
  commentstyle=\color{codegreen},
  keywordstyle=\color{magenta},
  numberstyle=\tiny\color{codegray},
  stringstyle=\color{codepurple},
  basicstyle=\ttfamily\footnotesize,
  breakatwhitespace=false,         
  breaklines=true,                 
  captionpos=b,                    
  keepspaces=true,                 
  numbers=left,                    
  numbersep=5pt,                  
  showspaces=false,                
  showstringspaces=false,
  showtabs=false,                  
  tabsize=2,
  xleftmargin=2em,
  numberstyle=\tiny\color{codegray},
}

\lstdefinelanguage{diff}{
  keywords={+, -, !},
  keywordstyle=\color{codegreen},
  morecomment=[l][\color{codegreen}]{+},
  morecomment=[l][\color{codered}]{-},
}
\lstdefinestyle{diffstyle}{
  language=diff,
  basicstyle=\ttfamily\footnotesize,
  breaklines=true,
  numbers=left, % Enable line numbering
  numbersep=5pt, % Set separation between numbers and code
}

\definecolor{darkgreen}{rgb}{0.0, 0.5, 0.0}
\newcommand\golnazsout[1]{\bgroup\markoverwith{\textcolor{darkgreen}{\rule[.5ex]{2pt}{0.8pt}}}\ULon{#1}}

\definecolor{mycommentcolor}{RGB}{100,100,255}  % choose your desired color

\newcommand{\code}[1]{\text{\lstinline[basicstyle=\ttfamily\small, language=Java]~#1~}}
\newcommand{\coder}[1]{\text{\lstinline[basicstyle=\ttfamily\small]~#1~}}

\usepackage{xspace}
\newcommand{\tool}{\textsc{PyTation}\xspace}
\newcommand{\comparedTool}{\textsc{Cosmic Ray}\xspace}

\usepackage[T1]{fontenc}
\usepackage{makecell}

\makeatletter
\newcommand\newtag[2]{#1\def\@currentlabel{#1}\label{#2}}
\makeatother

\newcommand{\RFA}{RemFuncArg\xspace}
\newcommand{\RCF}{RemConvFunc\xspace}
\newcommand{\REC}{RemElCont\xspace}
\newcommand{\DEC}{RemExpCond\xspace}
\newcommand{\CUA}{ChUsedAttr\xspace}
\newcommand{\DAA}{RemAttrAcc\xspace}
\newcommand{\RMC}{RemMetCall\xspace}

\newcommand{\RFAFull}{Remove Function Argument\xspace}
\newcommand{\RCFFull}{Remove Conversion Function\xspace}
\newcommand{\RECFull}{Remove Element From Container\xspace}
\newcommand{\DECFull}{Remove Expression from Condition\xspace}
\newcommand{\CUAFull}{Change Used Attribute\xspace}
\newcommand{\DAAFull}{Remove Attribute Access\xspace}
\newcommand{\RMCFull}{Remove Method Call\xspace}
\newcommand{\AvgEquivPerPrj}{1.61\%\xspace}

\newcommand{\AvgEquivRange}{0--4.74\%\xspace}

\newcommand{\AvgHeuristicDetected}{3.67\%\xspace}

\newcommand{\AVGTime}{$\sim$66.76 minutes\xspace}
\newcommand{\MEDTime}{$\sim$5.55 minutes\xspace}

\newcommand{\AVGMutationTime}{$\sim$63.09 minutes\xspace}
\newcommand{\AVGStaticTime}{$\sim$0.09 seconds\xspace}
\newcommand{\AVGDynamicTime}{$\sim$211.47 seconds\xspace}
\newcommand{\AVGIdentificationTime}{$\sim$211.56 seconds\xspace}
\newcommand{\AVGPostProcessTime}{$\sim$1.41 microseconds\xspace}
\newcommand{\MaxTime}{$\sim$462.93 minutes\xspace}
\newcommand{\MaxTimePrj}{drf\xspace}
\newcommand{\MinTime}{$\sim$0.35 minutes\xspace}
\newcommand{\MinTimePrj}{python-patterns\xspace}

\newcommand{\infobox}[1]{
    \begin{tcolorbox}[colback=white, colframe=gray, boxrule=0.3mm, width=\columnwidth, arc=0mm,left=1mm,
        right=1mm,
        top=1mm,
        bottom=1mm]
      \normalfont\selectfont
      {#1}
    \end{tcolorbox}
}

\newif\ifannotated%
\annotatedtrue% % Set this to \annotatedfalse for a clean version
\usepackage{todonotes}

\usepackage{amsthm}
\theoremstyle{definition} 
\newtheorem{definition}{Definition} 

\usepackage{tabularx}

\copyrightyear{2026}
\acmYear{2026}
\setcopyright{cc}
\setcctype{by}
\acmConference[ICSE '26]{2026 IEEE/ACM 48th International Conference on Software Engineering}{April 12--18, 2026}{Rio de Janeiro, Brazil}
\acmBooktitle{2026 IEEE/ACM 48th International Conference on Software Engineering (ICSE '26), April 12--18, 2026, Rio de Janeiro, Brazil}
\acmPrice{}
\acmDOI{10.1145/3744916.3787804}
\acmISBN{979-8-4007-2025-3/2026/04}

\begin{document}

\title{Hybrid Fault-Driven Mutation Testing for Python}

\author{Saba Alimadadi}
\affiliation{%
  \institution{Simon Fraser University}
  \city{Burnaby}
  \state{BC}
  \country{Canada}}
\email{saba@sfu.ca}

\author{Golnaz Gharachorlu}
\affiliation{%
 \institution{University of Ottawa}
 \city{Ottawa}
 \state{ON}
 \country{Canada}}
\email{ggharach@uottawa.ca}
\renewcommand{\shortauthors}{Alimadadi and Gharachorlu}

\begin{abstract}

Mutation testing is an effective technique for assessing the effectiveness of test suites
by systematically injecting artificial faults into programs.
However, existing mutation testing techniques fall short in capturing many types of
common faults in dynamically-typed languages like Python.
In this paper, we introduce a novel set of seven mutation operators that are inspired
by prevalent anti-patterns in Python programs, designed to complement the existing
general-purpose operators and broaden the spectrum of simulated faults.
We propose a mutation testing technique that utilizes a hybrid of static and dynamic analyses to mutate Python programs based on these operators while minimizing equivalent mutants.
We implement our approach in a tool called \tool and evaluate it on 13 open-source
Python applications. 
Our results show that \tool generates mutants that complement those from general-purpose tools, exhibiting distinct behaviour under test execution and uncovering inadequacies in high-coverage test suites. We further demonstrate that \tool produces a high proportion of unique mutants, a low cross-kill rate, and a low test overlap ratio relative to baseline tools, highlighting its novel fault model. \tool also incurs few equivalent mutants, aided by dynamic analysis heuristics.
\end{abstract}

\maketitle

\section{Introduction}
\label{sec:introduction}
Software testing is a crucial phase in software development to ensure code quality.
Mutation testing is a widely-studied technique for assessing the adequacy of a test suite~\cite{Jia11, Just14, Offutt2001, Papadakis19, Petrovic21, Sanchez24}.
It is a fault-based approach that injects artificial faults, i.e. \textit{mutations}, into the code using predefined \textit{mutation operators}.
A test suite is then exercised on the mutated code (\textit{mutant}) to determine its effectiveness in detecting these faults.
A mutant is considered \textit{killed} if at least one test fails and \textit{survives} otherwise.
Surviving (\textit{alive}) mutants reveal the inadequacies of tests in detecting the injected faults.

Mutation testing quantifies the effectiveness of the test suite using the \textit{mutation score}, the ratio of killed mutants to the total number of mutants.
A high mutation score suggests a stronger test suite.
However, a key challenge in mutation testing is the presence of \textit{equivalent mutants}, which modify the code syntactically but do not affect its behaviour, making them undetectable by tests~\cite{Offutt2001}.
The presence of equivalent mutants can skew the mutation score and increase the cost of analysis.
	
Various mutation testing techniques have been developed for different programming languages, including Java~\cite{PIT, Offutt2004, Javalanche}, C/C++~\cite{Delgado17MuCPP, Derezinska11}, and JavaScript~\cite{Mirshokraie13}.
However, Python's dynamic typing, flexible argument passing, and runtime attribute resolution create fault scenarios that are distinct from those in statically-typed languages.
For example, Python allows removing required function arguments without triggering a compile-time error or accessing arbitrary attributes via reflection, making it easier for subtle faults to go unnoticed by general-purpose mutation operators.
Common anti-patterns such as overly permissive function signatures, silent type coercions, and incomplete conditionals are frequent in Python codebases, yet difficult to catch with standard mutation techniques.
%Python's dynamic nature and the use of Pythonic idioms, such as list comprehensions and built-in functions with high-level abstractions that conceal much of the implementation details, introduce unique fault patterns. Despite being common, these fault patterns are not effectively targeted by existing mutation operators.
This has led to the development of Python mutation testing tools in practice~\cite{Cosmicray, Derezinska14}.\footnote{\url{https://mutmut.readthedocs.io/en/latest/}}\footnote{\url{}https://github.com/mutpy/mutpy}
While useful, these tools primarily apply general-purpose, syntax-based mutation operators and do not specifically target faults arising from Python's dynamic features or commonly occurring anti-patterns. These existing tools simulate faults at the level of syntax and structure, such as deleting statements or altering arithmetic operators, but fail to capture the kinds of semantic or runtime-specific issues prevalent in Python codebases.

Recent studies have analyzed frequent fault patterns in Python applications, highlighting limitations in current testing techniques and the need for more targeted approaches~\cite{DBLP:journals/corr/abs-2401-15481, PySStuBs}.
Despite the prevalence of these faults and Python's widespread adoption,\footnote{\url{https://survey.stackoverflow.co/2023/}}\footnote{\url{https://madnight.github.io/githut/}}\footnote{\url{https://www.tiobe.com/tiobe-index/}} there is currently no mutation testing technique that simultaneously targets Python-specific fault models and incorporates dynamic analysis to simulate and validate fault behaviors at runtime.

In this paper, we introduce a novel fault-driven mutation testing technique for Python that leverages both static and dynamic analyses to identify and simulate Python-specific fault patterns.
First, we discuss a set of common Python anti-patterns and introduce seven new mutation operators that mutate the code to artificially create these anti-patterns.
Unlike prior techniques that operate purely statically, our dynamic instrumentation enables us to simulate faults that manifest only under specific runtime conditions, such as the removal of defaulted arguments or dynamic attribute lookups.
%\annotate{meta comment 2: emphasize empirical basis of operators}%
%\annotate{AC1}%
To ensure practical relevance, our mutation operators are derived from prevalent, Python-specific bug/fix patterns categorized in a recent empirical study of 1{,}000 projects~\cite{PySStuBs}.
Then, we propose a technique based on a hybrid of static and dynamic analyses that mutates the code using our operators and applies a set of heuristics to reduce the number of equivalent mutants by eliminating those whose execution trace and state remain unchanged post-mutation.
We implement our approach in a tool called \tool and evaluate it using 13 open-source Python applications.
We further analyze how the mutants generated by \tool differ from those produced by a state-of-the-art Python mutation testing tool.

In summary, our contributions are as follows.

\begin{itemize}

\item A novel set of seven mutation operators derived from common Python anti-patterns, which broaden the mutation space and complement existing general-purpose mutation operators.
    
\item A hybrid static and dynamic mutation testing technique for generating these mutants, integrated with heuristics to reduce the number of equivalent mutants by observing runtime behaviour.
    
\item An open-source implementation of our approach in a tool called \tool,
along with all experimental data and scripts for reproducing our results,
available at \url{https://github.com/SEatSFU/pytation}.

\item An empirical evaluation of \tool on 13 real-world Python applications, demonstrating its ability to uncover inadequacies in Python tests. We also contextualize our technique by examining how it complements a state-of-the-art Python mutation testing tool.
    
\end{itemize}

Our results show that \tool identifies gaps in Python test suites by injecting realistic, Python-specific faults. It reveals inadequacies even in high-coverage code and produces a large proportion of unique mutants. While maintaining a low equivalent mutant rate, it complements general-purpose tools like \comparedTool by uncovering distinct behavioural faults and requiring different test coverage.
Rather than replacing existing mutation testing methods, our technique augments them by exposing faults that general-purpose operators often miss, yielding deeper insight into test effectiveness.

\section{Mutation Operators}
\label{Sec:mutation_operators}

We introduce a set of seven Python-specific mutation operators, each designed to simulate the effects of a recurring anti-pattern frequently observed in real-world Python programs (summarized in~\autoref{mo_table}).
These anti-patterns represent recurring coding practices that reduce code reliability and are especially prevalent in dynamically-typed, idiomatic Python code.
They are empirically grounded in a large-scale study~\cite{PySStuBs}, which analyzed over 1{,}000 open-source projects and identified common single-statement bug/fix patterns.
From these, we selected frequent Python-specific categories and designed operators that revert fixed code to its faulty form, focusing on faults characteristic of Python’s dynamic semantics while omitting generic mutation types already covered by existing tools.
To evaluate whether existing test suites are equipped to detect such issues, we simulate their effects through carefully designed mutations.
Each operator transforms a code pattern into a less robust or misleading form, mimicking a known anti-pattern while preserving syntactic validity.

%%%%%%%%%%%%%%%%%%%%% MUTATION OPERATOR 1 %%%%%%%%%%%%%%%%%%%%%

\subsection{\RFAFull (\RFA)}

Python functions can accept a flexible number of arguments via default parameters, \code{*args}, or \code{**kwargs}, making argument-related mistakes common~\cite{PySStuBs}.
As an example,\footnote{All examples in this section are simplified.} consider the \code{get_encoder()} function from GPT-2,\footnote{\url{https://github.com/openai/gpt-2/}} a large language model developed by OpenAI.
This function reads a vocabulary file (lines~\ref{line:encoding_bug}--\ref{line:read_encoding_data}) and initializes an \code{Encoder} object for tokenizing model inputs (line~\ref{line:encoder_initialization}).

\begin{lstlisting}[language=diff, escapechar=|]
def get_encoder(): |\label{line:encoding_bug_start}|
    ...
  - with open('data.bpe', 'r') as f: |\label{line:encoding_bug}|
  + with open('data.bpe', 'r', encoding='utf-8') as f: |\label{line:encoding_fix}|
        encoding_data = f.read() |\label{line:read_encoding_data}|
    ...
    return Encoder(data=encoding_data) |\label{line:encoder_initialization}|
\end{lstlisting}

A bug report revealed that this function crashed on Windows\footnote{\url{https://github.com/openai/gpt-2/pull/28}} because the file was opened without specifying an encoding. 
Without an explicit
encoding argument, the optional third argument (line~\ref{line:encoding_bug}), the application used the default file encoding specified by the operating system.
On non-Windows systems, the default was \code{utf-8}, but Windows defaulted to \code{cp1252}, causing decoding errors. 
The fix involved adding the explicit \code{utf-8} encoding (line~\ref{line:encoding_fix}).

The \RFAFull (\textit{\RFA}) operator simulates such bugs by removing one optional argument at a time from function calls that accept variable arguments, generating multiple mutants per call (Table~\ref{mo_table}, row~\ref{row:RFA}). 
This examines whether the suite can catch omissions that may silently change functionality, thereby assessing the robustness of the program.

\begin{table}[t]
\centering
\scriptsize
\setlength{\tabcolsep}{4pt}
\renewcommand{\arraystretch}{1.1}

\caption{Our proposed set of mutation operators} % and their example transformations.}
\label{mo_table}
\resizebox{\columnwidth}{!}{%
\begin{tabular}{@{}l l l@{}}
    \toprule
    \textbf{Mutation Operator} &
    \textbf{Original} &
    \textbf{Mutant} \\
    \midrule

    \newtag{1}{row:RFA}. \RFAFull (\RFA) &
    \(func(\dots, a_n=v_n)\) &
    \(func(\dots)\) \\

    \newtag{2}{row:RCF}. \RCFFull (\RCF) &
    \(conv\_func(x)\) &
    \(x\) \\

    \newtag{3}{row:REI}. \RECFull (\REC) &
    \([\dots, e_i, \dots]\) &
    \([\dots, e_{i-1}, e_{i+1}, \dots]\) \\

    \newtag{4}{row:DEI}. \DECFull (\DEC) &
    \(if \,\,cond1\,\, and\,\, cond2:\) &
    \(if\,\, cond1:\) \\

    \newtag{5}{row:CUA}. \CUAFull (\CUA) &
    \(obj.attr\) &
    \(obj.other\_attr\) \\

    \newtag{6}{row:DAA}. \DAAFull (\DAA) &
    \texttt{obj.attr} &
    \texttt{obj} \\

    \newtag{7}{row:RMC}. \RMCFull (\RMC) &
    \texttt{obj.method()} &
    \texttt{obj} \\

    \bottomrule
\end{tabular}
}
%\vspace{-10pt}
\end{table}

%%%%%%%%%%%%%%%%%%%%% MUTATION OPERATOR 2 %%%%%%%%%%%%%%%%%%%%%

\subsection{\RCFFull (\RCF)}
\label{Sec:MO2}

Python's dynamic typing enables flexible type conversions, but it also prevents static type checking, increasing the risk of runtime type errors~\cite{PyTER}.
Consider the following example from Home Assistant,\footnote{\url{https://github.com/home-assistant/core}} a popular home automation platform.
The \code{splunk_event_listener()} function logs system events, such as sensor readings, % or user preference changes
into Splunk, an external database. It builds a JSON-formatted request body from the event (lines~\ref{line:json_start}--\ref{line:json_end}) and sends it to the database (line~\ref{line:DCF_send_state}).

\begin{lstlisting}[language=diff, escapechar=|]
def splunk_event_listener(event): |\label{line:DCF_func_def}|
  ...
  request_body = [{ |\label{line:json_start}|
    ...,
  - 'attributes': event.attributes, |\label{line:DCF_error}|
  + 'attributes': dict(event.attributes) |\label{line:DCF_fix}|
  }] |\label{line:json_end}|
  post_request(convert_to_json(request_body)) |\label{line:DCF_send_state}|
\end{lstlisting}

A reported bug\footnote{\url{https://github.com/home-assistant/core/issues/1252}} exposed a \code{TypeError} during JSON conversion (line~\ref{line:DCF_send_state}), stemmed from the \code{attributes} field (line~\ref{line:DCF_error}), which held a \code{MappingProxy}, a non-serializable type.
Because Python lacks compile-time type checks, this error surfaced only at runtime.
The fix involved explicitly converting the field to a \code{dict} before serialization (line~\ref{line:DCF_fix}).

%\annotate{meta comment 5: clarify meaning of explicit conversions and heuristic (R1.C4, R2.C5)}%
The \RCFFull (\textit{\RCF},~\autoref{mo_table}, row~\ref{row:RCF}) removes explicit type conversions
(e.g., \texttt{int()}, \texttt{str()}, \texttt{list()}) to expose reliance on implicit coercions like \texttt{\_\_str\_\_}. 
It models bugs where developers omit necessary conversions. 
To prevent equivalent mutants, \tool\ skips mutations when the runtime type of the argument already matches the conversion’s return type.

%%%%%%%%%%%%%%%%%%%%% MUTATION OPERATOR 3 %%%%%%%%%%%%%%%%%%%%%

\subsection{\RECFull (\REC)}

Python containers (e.g., lists and dictionaries) are widely used for managing collections of data.
Due to Python’s dynamic nature, developers may incorrectly assume the structure or size of containers, often accessing or unpacking elements that do not exist~\cite{PySStuBs}.
%Due to the language's dynamism, it is common for developers to attempt to access indices, keys, and elements that do not exist in the containers based on incorrect assumptions about their contents and structure~\cite{PySStuBs}.

Consider the \code{DGLGraph} class from the Deep Graph Library (DGL),\footnote{\url{https://github.com/dmlc/dgl}} a scalable Python framework for deep learning on graphs.
This class provides a flexible and efficient graph representation that users can customize.
The method \code{batch_update_edge()} updates all edges in the graph.
It first retrieves the edges via \code{get_edges()} (line~\ref{line:DEI_bug}) and then applies an \code{updater_function()} to them (line~\ref{line:DEI_update}).

\begin{lstlisting}[language=diff, escapechar=|]
class DGLGraph(): |\label{line:DEI_class_def}|
  def batch_update_edge(self, updater_function): |\label{line:DEI_batch_update}|
      - u, v = self.get_edges() |\label{line:DEI_bug}|
      + u, v, _ = self.get_edges() |\label{line:DEI_fix}|
        new_edges = updater_function(u, v) |\label{line:DEI_update}|
        ... |\label{line:DEI_batch_update_end}|
  def get_edges(self): |\label{line:DEI_get_edges}|
    # Retrieve all edges
    return (src, dst, eid) |\label{line:DEI_get_edges_return}|
\end{lstlisting}

A user-reported crash\footnote{\url{https://github.com/dmlc/dgl/issues/81}} revealed that \code{get_edges()} returns a tuple of three lists (sources, destinations, and edge IDs) (line~\ref{line:DEI_get_edges_return}).
However, \code{batch_update_edge()} expected only two variables (source and destination vertices, line~\ref{line:DEI_bug}), causing a runtime \code{ValueError} due to unpacking mismatch.
The developers fixed the issue by introducing a third variable (line~\ref{line:DEI_fix}) using the underscore \code{\_}, a common Python convention for unused values.

The \RECFull (\textit{\REC}) operator simulates such structural mismatches by removing a randomly selected element from a container (Table~\ref{mo_table}, row~\ref{row:REI}).
This exposes test suites to faults caused by incorrect assumptions about container contents.

%%%%%%%%%%%%%%%%%%%%% MUTATION OPERATOR 4 %%%%%%%%%%%%%%%%%%%%%

\subsection{\DECFull (\DEC)}

%Conditional statements can be difficult to test. 

Compound conditional statements are commonly used to succinctly express multi-part logic.
However, they are also prone to subtle bugs when developers fail to test all subconditions effectively.
Such issues arise when a critical part of a compound condition is missing, misordered, or silently evaluates to an unexpected value (e.g., \code{None} or \code{False})~\cite{DBLP:journals/ese/PanKW09, 6747176}.
The following example is from Ansible, a popular IT automation tool.\footnote{\url{https://github.com/ansible/ansible}}
The \code{handle_download()} function downloads a file from a remote URL (\code{remote_file_url}) to a target directory (\code{destination}), optionally using a checksum to skip downloads of existing files (\code{remote_file_checksum}).
A checksum is a content-derived signature used to verify file integrity~\cite{DBLP:journals/tcom/Fletcher82, sivathanu2004enhancing}.
The function computes a checksum at the destination path (line~\ref{line:DEIF_checksum_calc}).
If the checksum is not \code{None} (line~\ref{line:DEIF_checksum_if}), it is compared to the one provided (line~\ref{line:DEIF_checksum_if_body}).
If they match (line~\ref{line:DEIF_bug}), the download is skipped as a duplicate file already exists at the destination (line~\ref{line:DEIF_skip}); otherwise, the file is downloaded (line~\ref{line:DEIF_download}).

\begin{lstlisting}[language=diff, escapechar=|]
def handle_download(remote_file_url, destination, remote_file_checksum=None): |\label{line:DEIF_handle_download}|
  checksum_mismatch = False |\label{line:DEIF_mismatch_defined}|
  ...
  checksum = calculate_checksum(f"{destination}{remote_file_name}") |\label{line:DEIF_checksum_calc}|
  if checksum: |\label{line:DEIF_checksum_if}|
    checksum_mismatch = (checksum |$\neq$| remote_file_checksum) |\label{line:DEIF_checksum_if_body}|

- if not checksum_mismatch: |\label{line:DEIF_bug}|
+ if checksum and not checksum_mismatch: |\label{line:DEIF_fix}|
      # Ansible skips the download, ending the automation |\label{line:DEIF_skip}|
  else:
      # Ansible proceeds to download the file |\label{line:DEIF_download}| |\label{line:DEIF_handle_download_end}|
\end{lstlisting}

A bug report\footnote{\url{https://github.com/ansible/ansible/issues/61978}} showed that if the destination file did not exist, the computed checksum was \code{None}, but the download was still skipped.
The conditional on line~\ref{line:DEIF_checksum_if} did not execute, and the \code{checksum_mismatch} variable (line~\ref{line:DEIF_mismatch_defined}) incorrectly remained \code{False}.
The fix updated the conditional to explicitly check for a non-\code{None} checksum (line~\ref{line:DEIF_fix}).

The \DECFull (\textit{\DEC}) operator simulates such oversights by removing one condition from compound conditionals (Table~\ref{mo_table}, row~\ref{row:DEI}).
This exposes tests to faults arising from missing logical branches.

\subsection{Object Oriented Mutation Operators}

Python’s object-oriented features, combined with dynamic typing, allow developers to access attributes or methods that do not exist, without triggering compile-time errors~\cite{PySStuBs}.
We introduce three mutation operators to simulate anti-patterns in object interaction.

%%%%%%%%%%%%%%%%%%%%% MUTATION OPERATOR 5 %%%%%%%%%%%%%%%%%%%%%

\subsubsection{\CUAFull (\CUA)}

In statically-typed languages, incorrect attribute access is caught at compile time. In contrast, Python’s dynamic typing delays such errors until runtime, making tests the primary defence against misused object attributes.
The following example from Modin,\footnote{\url{https://github.com/modin-project/modin}} a scalable data processing library, demonstrates the \code{sample()} function used for randomized sampling of data.
This function takes a \code{random\_state} value as an input argument, which acts as the seed for a random number generator. The input should either be a \code{RandomState} object from NumPy\footnote{\url{https://github.com/numpy/numpy}} or an integer from which one can be constructed.

\begin{lstlisting}[language=diff, escapechar=|]
def sample(random_state): |\label{line:CUA_def}|
  ...
  if isinstance(random_state, int): |\label{line:CUA_check_int}|
    random_num_gen = numpy.random.RandomState(random_state) |\label{line:CUA_convert}|
- elif isinstance(random_state, numpy.random.randomState): |\label{line:CUA_bug}|
+ elif isinstance(random_state, numpy.random.RandomState): |\label{line:CUA_fix}|
    random_num_gen = random_state |\label{line:CUA_end}|
\end{lstlisting}

A user reported an \code{AttributeError} upon invoking this function with a valid Numpy \code{RandomState}.\footnote{\url{https://github.com/modin-project/modin/issues/1692}}
The error was caused by a typo in the attribute name of the \code{numpy.random} module: \code{randomState} instead of the correct \code{RandomState} (line~\ref{line:CUA_bug}).
Since both names are valid identifiers, the mistake remained undetected until runtime.
The bug was resolved by correcting the attribute name (line~\ref{line:CUA_fix}).

The \CUAFull (\textit{\CUA}) mutation operator simulates this class of errors by replacing an accessed attribute with another randomly-selected attribute of the same object (row~\ref{row:CUA} of Table~\ref{mo_table}).
This operator captures the fragility of attribute-based access in dynamically-typed languages and highlights insufficient testing of object structure assumptions.

%%%%%%%%%%%%%%%%%%%%% MUTATION OPERATOR 6 %%%%%%%%%%%%%%%%%%%%%

\subsubsection{\DAAFull (\DAA)} 

Extending the previous anti-pattern, Python code may mistakenly use an object itself where a specific attribute should have been accessed. Omitting an attribute leads to runtime errors that would otherwise be caught statically in typed languages.

The following example is from \textit{Electrum}, a popular Bitcoin wallet.\footnote{\url{https://github.com/spesmilo/electrum}}
It shows the \code{HistoryList} class, responsible for generating and displaying the user's transaction history.
\code{HistoryList} includes a \code{history_model} attribute, an instance of \code{HistoryModel}, which manages the stored transaction data (line~\ref{line:DAA_hm_attr}).
To export this data, the \code{do_export_history()} method invokes \code{get_full_history()} with a \code{domain} parameter that scopes the exported data (line~\ref{line:DAA_bug}).

\begin{lstlisting}[language=diff, escapechar=|]
class HistoryList: |\label{line:DAA_hl_class}|
  def __init__(self, model, wallet):
    self.history_model = model |\label{line:DAA_hm_attr}|
    self.wallet = wallet
    ...

  def do_export_history(self): |\label{line:DAA_method}|
  - hist = self.wallet.get_full_history(domain=self.get_domain(), ...) |\label{line:DAA_bug}|
  + hist = self.wallet.get_full_history(domain=self.history_model.get_domain(), ...) |\label{line:DAA_fix}|
    ...
    export_to_file(hist) |\label{line:DAA_file}|
\end{lstlisting}

This feature caused crashes reported via the application's crash system.\footnote{\url{https://github.com/spesmilo/electrum/issues/4948}}
The error stemmed from line~\ref{line:DAA_bug}, where \code{self.get_domain()} was called directly, omitting the required \code{history_model} attribute.
\code{HistoryList} lacks such a method, causing an \code{AttributeError}. The bug was fixed by accessing \code{get_domain()} through \code{history_model} as intended (line~\ref{line:DAA_fix}).

The \DAAFull (\textit{\DAA}) mutation operator simulates this class of faults by deleting an attribute access and replacing it with the base object (e.g., \code{self}), mimicking real-world cases where attribute references are omitted (row~\ref{row:DAA} of Table~\ref{mo_table}).

%%%%%%%%%%%%%%%%%%%%% MUTATION OPERATOR 7 %%%%%%%%%%%%%%%%%%%%%
        
\subsubsection{\RMCFull (\RMC)}

Another frequent object-related anti-pattern in Python involves mistakenly omitting method calls. Due to Python’s dynamic typing, such omissions are not caught at compile time and can silently fail at runtime.
The following example is from \textit{Read the Docs},\footnote{\url{https://github.com/readthedocs/readthedocs.org}} a widely-used platform for hosting documentation, which allows projects to have multiple versions. Versions can be marked as inactive (unavailable) or private (restricted to specific users).

One feature shows an alert to users viewing outdated documentation, offering a link to the latest version. Clicking the alert redirects users to that version.
This feature uses the \code{get_version_compare_data()} function to retrieve the latest version of a project.
It does so by passing all active versions of the project to the \code{highest_version()} function (lines~\ref{line:RMC_highest_ver} and~\ref{line:RMC_bug}), which returns the most recent version.

\begin{lstlisting}[language=diff, escapechar=|]
def get_version_compare_data(project): |\label{line:RMC_get_version_def}|
    highest_ver_obj = highest_version( |\label{line:RMC_highest_ver}|
       - project.versions.filter(active=True) |\label{line:RMC_bug}|
       + project.versions.public().filter(active=True) |\label{line:RMC_fix}|
    ) |\label{line:RMC_end}|
\end{lstlisting}

However, the system began redirecting users to private (non-visible) pages when the latest version was not public.\footnote{\url{https://github.com/readthedocs/readthedocs.org/pull/1794}}
The issue occurred because private versions were not filtered out from \code{project.versions}.
The developers fixed this by calling the \coder{public()} method on \code{project.versions}, restricting the list to public versions only (line~\ref{line:RMC_fix}).

The \RMCFull (\textit{\RMC}) mutation operator simulates this class of errors by removing method calls and replacing them with the base object.
This simulates scenarios where developers forget to invoke methods that return the appropriate filtered or transformed data (row~\ref{row:RMC} of Table~\ref{mo_table}).

%\annotate{meta comment 4: address concern on operator specificity (R3.C2)}%
Although most operators focus on Python-specific constructs, their underlying fault patterns (e.g., argument removal, condition simplification, attribute misuse) are broadly recurring across languages. 
Their empirical grounding in real Python projects further supports their prevalence and potential generality beyond our evaluated benchmarks~\cite{PySStuBs}.

\section{Approach}
\label{Sec:approach}

We present \tool's mutation testing framework for Python, which is designed to simulate dynamic-language-specific fault patterns and assess test adequacy.
The process begins by identifying \textit{mutation candidates}, i.e., code locations where Python-specific anti-patterns may arise, based on the operators introduced in~\autoref{Sec:mutation_operators}.
To achieve this, \tool employs a novel hybrid analysis that combines static Abstract Syntax Tree (AST) analysis with dynamic analysis, enabling both broad candidate discovery and precise mutation targeting.

Static analysis is used to detect mutation candidates that can be determined purely syntactically, such as compound conditions or list literals targeted by \REC and \DEC.
Dynamic analysis is essential for identifying context-sensitive candidates, such as optional function arguments, runtime attribute accesses, or dynamically-constructed method calls. These require concrete runtime values and object structures not evident statically.
In addition, runtime traces enable \tool to apply a set of heuristics to prune mutation candidates that are unlikely to yield semantically meaningful changes. These heuristics help eliminate likely-equivalent mutants, and discard mutations in code not exercised by the test suite.

\begin{figure}[t]
    \centering
    \includegraphics[width=\columnwidth]{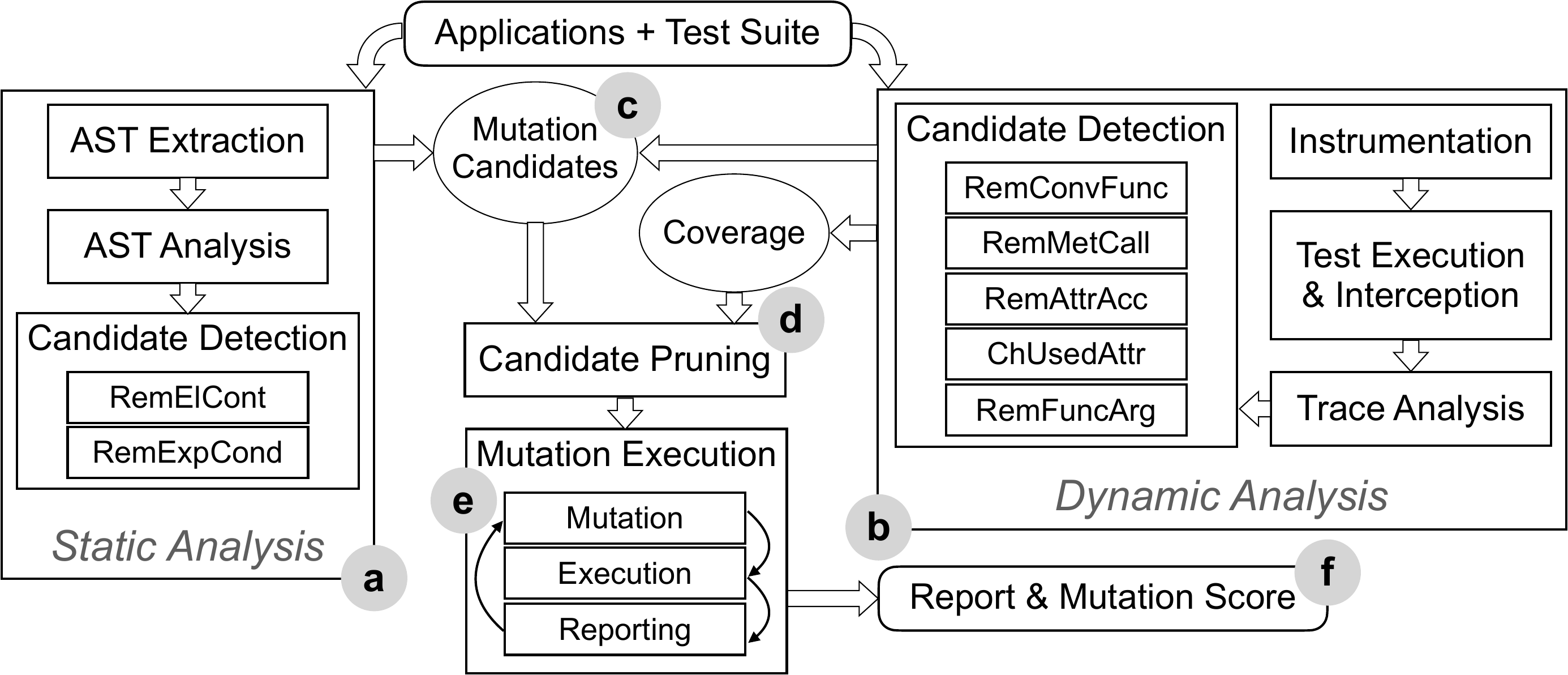}
    \caption{Overview of \tool. The approach combines static (a) and dynamic (b) analyses to identify mutation candidates, prunes unreachable locations, and generates mutants (c) for mutation score computation (d).}
    \label{fig:approach_draft}
    %\saba{need to draw a better and more professional diagram}
%\vspace{-14pt}
\end{figure}

Each valid mutation is then injected into the original program, the test suite is executed, and the outcome (killed or survived) is recorded.
This process is repeated across all mutation candidates, generating a fine-grained view of test effectiveness across a diverse range of dynamic fault models.
\autoref{fig:approach_draft} illustrates the overall architecture and workflow of our hybrid mutation analysis.

\subsection{Identifying Mutation Candidates}

\tool identifies mutation candidates through a hybrid analysis combining static AST traversal and dynamic runtime interception, tailored to the anti-patterns defined in~\autoref{Sec:mutation_operators}.
Each candidate (\autoref{mo_table}, \textit{Original} column) corresponds to an instance of a Pythonic programming construct that is either syntactically suspicious or contextually fragile, i.e., likely to exhibit faulty behaviour under minor perturbations. These constructs may incur recurring anti-patterns in dynamic Python code (see \autoref{mo_table}, \textit{Mutant} column).
The outcome of this phase is a structured list of mutation candidates (see~\autoref{fig:approach_draft}-c), where each candidate is paired with a specific operator and its source-level context, enabling targeted injection of faults.

Formally, we represent each mutation candidate as a tuple \(MC = \langle Label, LOC, Metadata \rangle\), where \textit{Label} identifies the operator to apply, \textit{LOC} denotes the precise code location (filename, line/column range) of the target AST node, and \textit{Metadata} encodes auxiliary information gathered during static or dynamic analysis (such as argument positions, available attributes, observed runtime types, or container elements) that guides the mutation logic.
\subsubsection{Static Analysis}

We use static analysis to identify mutation candidates for \RECFull (\REC) and \DECFull (\DEC), which target anti-patterns manifesting in purely syntactic code structures.
Specifically, container initializations and compound boolean conditions can be detected by traversing the program's abstract syntax tree (AST), without requiring runtime information.
Our approach first parses the source code into its AST representation and then traverses the tree to detect syntactic structures matching the patterns targeted by these operators.
This static candidate detection phase is shown in~\autoref{fig:approach_draft}-a.
Each AST node encodes a syntactic construct (e.g., a list literal or boolean condition) along with its attributes and substructure.
Below, we describe how our analysis statically detects candidates for the two operators.

Our analysis locates AST nodes corresponding to container literals, such as lists, tuples, sets, and dictionaries, that contain at least one element. These nodes represent points in the code where structural assumptions about container contents may lead to faults.
We add a new mutation candidate labeled \textit{\REC} for each such node, recording its precise source location.
The AST structure provides direct access to list, tuple, and set elements, while dictionary literals store keys and values in parallel lists, preserving their mapping.
We include in the candidate's metadata the element count and index mappings to enable later mutation generation.

\paragraph{\DEC}

To identify \DEC candidates, we locate AST nodes corresponding to conditional constructs, including \texttt{if} and \texttt{while} statements and expressions involving boolean connectives (e.g., \texttt{a = exp1 or exp2}).
When a condition contains multiple subclauses joined by \texttt{and} or \texttt{or}, the AST encodes them as a sequence of operands under a shared logical operator node.
Our analysis recursively decomposes such compound expressions to enumerate all boolean subcomponents.
If the overall condition comprises more than one boolean clause, it is marked as a mutation candidate. We add an entry labeled \textit{\DEC}, capturing the full expression and its source position.
\subsubsection{Dynamic Analysis}

As illustrated in~\autoref{fig:approach_draft}-b, our approach detects mutation candidates that cannot be reliably identified through static analysis alone, due to Python’s dynamic semantics.
To this end, we perform dynamic analysis by automatically instrumenting the target application and intercepting its execution under its existing test suite.
This yields execution traces that we analyze to identify candidates for the \RFAFull (\RFA), \RCFFull (\RCF), \CUAFull (\CUA), \DAAFull (\DAA), and \RMCFull (\RMC) operators, each of which targets anti-patterns that manifest dynamically.

We log function calls (including names and argument values) and object interactions, such as attribute and method accesses, capturing both the accessed member and its receiver object.
Candidates extracted during this phase are annotated with metadata derived from runtime behaviour.
To minimize unnecessary mutations and reduce equivalent mutants, we apply targeted heuristics during candidate extraction, such as type compatibility checks and object attribute filtering.
We describe below how dynamic analysis is used to detect mutation candidates for each of these five operators.

\paragraph{\RFA}
We analyze function calls by inspecting the relationship between provided arguments and function signatures.
A new mutation candidate labeled \textit{\RFA} is created if any of the following runtime conditions hold:
(i) A default parameter is explicitly passed at the call site,
(ii) The function uses \code{*args} and receives additional positional arguments, or
(iii) The function accepts \code{**kwargs} and is called with keyword arguments not explicitly declared.
Each such case is recorded as a candidate with its call-site location ($LOC$) and the removable argument’s index or name ($Metadata$).

\paragraph{\RCF}
For each call site, we check whether the callee is one of Python’s built-in conversion functions (e.g., \code{int}, \code{str}, \code{list}, \code{dict}, \code{tuple}, etc.).
To avoid producing equivalent mutants, we use a type-based heuristic: if the runtime type of the argument already matches the function’s return type, the mutation is skipped.
Otherwise, we record a \textit{\RCF} candidate with its source location and the function name.

\paragraph{\CUA}
%
%As a heuristic, we discard candidates where the object has only a single attribute, as no meaningful alternative exists.
We treat all attribute accesses as potential \textit{\CUA} candidates. To ensure mutation validity, we discard cases where the receiver object exposes only one attribute.
For valid candidates, we randomly select an alternate accessible attribute from the same object instance and store it in $Metadata$.
We also exclude reserved Python special (dunder) attributes (e.g., \code{__name__}, \code{__doc__}) to avoid interfering with interpreter behavior.
Each valid \textit{\CUA} candidate includes its location and the selected alternative attribute.

\paragraph{\DAA}
An attribute access is also considered a candidate for \textit{\DAA}, which simulates faults involving the omission of a required attribute access.
These candidates are labeled \textit{\DAA} and stored with their access site location.

\paragraph{\RMC}
We mark method calls, i.e., invocations of bound functions on object instances, as candidates for the \textit{\RMC} operator, which simulates omission of such invocations. Each call is recorded with the \textit{\RMC} label and its location.

\subsubsection{Pruning Uncovered Candidates}
Mutation candidates not exercised by any test case cannot meaningfully contribute to mutation testing, as their injected faults remain unobserved during execution.
Therefore, our approach uses  coverage information to discard statically-identified candidates that are not reached by any test (\autoref{fig:approach_draft}-d).
The result is a reduced and more meaningful set of test-reachable, syntactically-valid candidates prepared for mutation.
%
%\annotate{meta comment 6: clarify purpose of dynamic analysis and test-suite dependence (R3.Q2)}%
\tool\ uses dynamic analysis to insert mutants only at locations exercised by the given test suite, ensuring executability without increasing killability: execution alone is insufficient without suitable oracles. 
This design avoids unreachable mutants that would always survive and aligns with prior coverage-guided mutation approaches~\cite{Petrovic21,Mirshokraie13,Gligoric11} and  tools (Mutatest\footnote{\url{https://mutatest.readthedocs.io/en/latest/index.html}} and Poodle.\footnote{\url{https://poodle.readthedocs.io/en/latest/}})

\subsection{Mutating Process}
\label{subsec:mutating-process}

In the second phase of our approach (\autoref{fig:approach_draft}-e), we apply each mutation individually to the original program, execute the test suite, and record the outcome.
We log the outcome of each test (pass/fail) for every mutated version.
If at least one test fails, the mutant is considered \textit{killed}; otherwise, it has \textit{survived} and is considered \textit{alive}.
For killed mutants, we also record the identity of the failing tests.

After evaluating a mutant, we revert the program to its original form before applying the next mutation.
After all mutations have been tested, we compute the overall mutation score and generate a report summarizing killed and alive mutants  (\autoref{fig:approach_draft}-f).

\subsubsection{Mutating the Code}

To apply a mutation, our approach retrieves the mutation candidate’s AST node using the $LOC$ field.
The $Label$ field specifies the mutation operator, which dictates the transformation logic and the applicable AST node type.
The appropriate transformation is applied to the AST node, after which the mutated tree is serialized back to source code.
The resulting mutant is tested immediately using the application’s test suite.
Each operator applies a targeted transformation, grounded in Python’s semantics and corresponding anti-patterns:

\begin{itemize}
    \item \textit{\REC} removes a randomly-selected element from a container literal (e.g., list, tuple, set, or dictionary), simulating errors in collection initialization.
    
    \item \textit{\DEC} removes one sub-condition from compound boolean expressions. % (e.g., \code{a and b}), simulating incomplete guard logic.

    \item \textit{\RFA} removes an optional argument from a function call, simulating errors due to omitted defaults or extra arguments.

    \item \textit{\RCF} replaces a type conversion function call with its raw input, exposing latent type mismatches.

    \item \textit{\CUA} substitutes one accessed attribute with a different attribute of the same object, simulating attribute selection mistakes.

    \item \textit{\DAA} deletes the attribute access altogether, replacing \code{obj.attr} with \code{obj}, capturing bugs from forgotten attribute use.

    \item \textit{\RMC} omits a method call, modeling errors where behaviour-defining methods are mistakenly omitted.
    
\end{itemize}

\subsubsection{Generating the Mutation Report}

Once each mutant is tested, we determine whether it was killed or survived.
A detailed report is generated, recording the outcome and listing the specific test cases that killed each mutant.
From these results, we compute the \textit{mutation score}, which quantifies test suite effectiveness.
The final report summarizes the mutant-level results and the aggregate mutation score, offering insights into the fault detection capabilities of the test suite.

\section{Implementation}
\label{sec:implementation}

We implemented our approach in an open-source Python tool called \tool.\footnote{\url{https://github.com/SEatSFU/pytation}}
For static analysis and mutation, we used Python’s \code{ast} module to parse code, apply operator-specific transformations via \code{NodeTransformer}, and regenerate executable source.
Dynamic analysis was implemented using DynaPyt~\cite{DynaPyt} with custom instrumentation. We extracted runtime metadata using the \code{inspect} module.
\code{pytest}\footnote{\url{https://docs.pytest.org/en/latest/}} for test execution and \code{coverage}\footnote{\url{https://coverage.readthedocs.io/en/7.4.1/}} to guide mutation pruning and track test-to-mutant mappings.
\section{Evaluation}
\label{sec:evaluation}

We evaluated \tool by applying it to 13 real-world Python applications, measuring mutation scores and analyzing the distribution of killed and surviving mutants.
We then analyzed the relationship between \tool and \comparedTool, a widely-used state-of-the-art Python mutation testing tool~\cite{Cosmicray}, by applying both to the same benchmark suite.
This allowed us to determine whether our operators injected faults not detected by test suites that otherwise performed well under standard mutation operators.
We aimed to highlight the complementary benefits of \tool, specifically, its ability to generate mutants that simulate distinct fault behaviours and expose gaps in existing test suites.
To this end, we studied the behavioural interplay between mutants from both tools, measuring cross-kill rates and test overlap ratios to quantify fault coupling and divergence.
Finally, we evaluated the runtime overhead of \tool to assess its practicality. % and suitability for real-world use.
In particular, we conducted three experiments to answer the following research questions.

\begin{itemize}

    \item \textbf{RQ1:} How effective is \tool at revealing inadequacies in existing Python test suites, and how many equivalent mutants does it produce?

    \item \textbf{RQ2:} How do the mutants generated by \tool complement those from existing tools like \comparedTool in terms of diversity, detectability, and test interaction?

    \item \textbf{RQ3:} What is the performance overhead of using \tool relative to baseline test execution?

\end{itemize}

\subsection{Experimental Setup}
\label{sec:experimental_setup}

\subsubsection{Experimental Subjects}

We evaluated \tool on 13 open-source Python projects from GitHub, listed in the first column of \autoref{tab:benchmarks}. All projects were under active development or maintenance and included automated \texttt{pytest}-based test suites.
These benchmarks span diverse domains, such as image processing, scientific computing, and web frameworks, and vary in size, complexity, and architecture. \autoref{tab:benchmarks} (columns 2–5) reports each project's lines of code (LOC), number of functions, test cases, and test coverage.
Project sizes range from 700 to 20K LOC (mean: 7.1K), achieving line coverage between 74\% and 99\% (mean: 93\%).

\subsubsection{Experiment 1}
\label{design:rq1}

To address \textbf{RQ1}, we ran \tool on all benchmark applications to evaluate how effectively their existing test suites detect faults introduced by our mutation operators.
We began by identifying all mutation candidates for each project.
For each candidate, we applied the corresponding operator to generate a single mutant, which we then executed against the application's test suite.
We recorded whether each mutant was \textit{killed} %(i.e., at least one test failed) 
or \textit{survived}, %(i.e., all tests passed), 
and computed the mutation score as the ratio of killed to total mutants.
We removed syntactically- and runtime-invalid mutants to avoid inflating scores. 
Invalids included mutants that could not be parsed, executed, or instrumented correctly.

A mutant is considered equivalent if its behaviour is indistinguishable from the original for all inputs. 
Because exact detection is undecidable, we follow common practice~\cite{offutt1997automatically, 10.1145/2568225.2568265, petrovic2018industrial, 10.1145/3650212.3680310} 
and approximate it through manual inspection of a sample of surviving mutants.
To manage the large number of candidates, we adopted selective mutation testing~\cite{10.1109/ASE.2013.6693070}.
Two independent non-author examiners reviewed 20\% of RQ1 mutants,
ensuring at least 10 mutants per operator per project, if possible.
They checked data and control dependencies (assignments, calls, returns, exceptions, I/O) to assess behavioural divergence.
We further evaluated the impact of our heuristics in reducing the number of equivalent mutants generated by \tool.

%%%%%%%%%%%%%%%%%%%%%%%%%%%%%%%%%%%%%%%%%%%%%%%%%%%%%%%%%%%%%%%%%%%%%%%%

\begin{table}[t]
{\centering
    \scriptsize
 \caption{General characteristics of the benchmarks.}
    \label{tab:benchmarks}
 % \setlength{\tabcolsep}{5pt}
 %    \renewcommand{\arraystretch}{1}
% \resizebox{0.48\textwidth}{!}{
    \begin{tabular*}{\columnwidth}{@{}l@{\extracolsep{\fill}}@{}rrrr@{}}
    \toprule

    \textbf{Project} & \textbf{LOC} & \textbf{\#Functions} & \textbf{\#Tests} & \textbf{Line Coverage\%} \\

    \midrule

    \newtag{1}{row:blinker}. \href{https://github.com/pallets-eco/blinker}{blinker}                  & 745   &   34   & 25   & 83     \\
    \newtag{2}{row:schedule}. \href{https://github.com/dbader/schedule}{schedule}                    & 945   &   57   & 81   & 98     \\
    \newtag{3}{row:pyjwt}. \href{https://github.com/jpadilla/pyjwt}{pyjwt}                           & 2,235  &   116  & 271  & 94     \\
    \newtag{4}{row:pyquery}. \href{https://github.com/gawel/pyquery}{pyquery}                        & 2,350  &   142  & 150  & 89     \\
    \newtag{5}{row:funcy}. \href{https://github.com/Suor/funcy}{funcy}                               & 2,369  &   262  & 202  & 96     \\
    \newtag{6}{row:wtforms}. \href{https://github.com/pallets-eco/wtforms}{wtforms}                  & 3,685  &   222  & 352  & 99     \\
    \newtag{7}{row:pythonpatterns}. \href{https://github.com/faif/python-patterns}{python\_patterns} & 3,779  &   352  & 65   & 92     \\
    \newtag{8}{row:marshmallow}. \href{https://github.com/marshmallow-code/marshmallow}{marshmallow} & 5,105  &   264  & 1,129 & 97     \\
%    \newtag{9}{row:pyvips}. \href{https://github.com/libvips/pyvips}{pyvips}                         & 6,180  &   249  & 53   & 74     \\
    \newtag{9}{row:structlog}. \href{https://github.com/hynek/structlog}{structlog}                 & 6,260  &   274  & 822  & 99     \\
    \newtag{10}{row:Flask}. \href{https://github.com/pallets/flask}{flask}                           & 8,979  &   391  & 1,223 & 95     \\
    \newtag{11}{row:graphene}. \href{https://github.com/graphql-python/graphene}{graphene}           & 12,255 &   764  & 468  & 96     \\
    % \newtag{12}{row:Supervisor}. \href{https://github.com/Supervisor/supervisor}{supervisor}         & 15,775 &   1,082 & 1,388 & 89     \\
    \newtag{12}{row:drf}. \href{https://github.com/encode/django-rest-framework}{django-rest-framework}                & 16,279 &   972  & 1,539 & 96     \\
    \newtag{13}{row:praw}. \href{https://github.com/praw-dev/praw}{praw}                             & 19,729 &   727  & 1,029 & 99     \\

    \midrule
    %           LOC     &  Functions &  Tests  &  Coverage        &  Mutation Score &  Time
    Mean &  6,517 &   352   &  566 &  95 \\

    \bottomrule
    \end{tabular*}
    }
   % }

%\saba{I think we should add branch coverage to this table and later analyze correlations between coverage metrics and mutation scores, etc. to see if we find something.}

%\vspace{-10pt}
\end{table}

\subsubsection{Experiment 2}

We analyzed how \tool's mutants differ from and complement those produced by \comparedTool 
We selected Cosmic Ray\footnote{\url{https://cosmic-ray.readthedocs.io/en/latest/}} as our baseline because it is among the most mature and actively-maintained Python mutation testing tools, offering the broadest operator set, parallel execution, and strong community adoption. 
It is widely used in SE research (e.g., \cite{Guerino2024} and practitioner resources, with over 460K recent downloads; far exceeding alternatives. 
Other tools (Mutmut, MutPy, Mutatest, Poodle) have narrower operator sets, limited scalability, or little recent maintenance (last commits 2--6 years ago; \cite{DBLP:conf/dsa/DialloCWL24}). 
Using Cosmic Ray ensured fairness, stability, and comparability with prior work.

\begin{table*}[t]
    \caption{The number of injected mutations (\#Mut) and the resulting mutation score (MS(\%)) per mutation operator and in total, for each application. Mutation operators include \RFA (RFA), \RCF (RCF), \REC (RElC), \DEC (RExC), \CUA (CUA), \DAA (RAA), and  \RMC (RMC), respectively. 
    % \saba{add links to repos for project names. especially since i'm reducing the length of some names.
    % (python\_patterns, marshmallow)}
    }
    \label{tab:mutations}

    \centering
    
    \scriptsize

	\resizebox{\textwidth}{!}{%
	
    \begin{tabular}{@{}l@{\extracolsep{\fill}}@{}rr@{}rr@{}rr@{}rr@{}rr@{}rr@{}rr@{}rr@{}}
    \toprule

    \multicolumn{1}{l}{} &

    \multicolumn{2}{c}{\textbf{RFA}} &
    \multicolumn{2}{c}{\textbf{RCF}} &
    \multicolumn{2}{c}{\textbf{RElC}} &
    \multicolumn{2}{c}{\textbf{RExC}} &
    \multicolumn{2}{c}{\textbf{CUA}} &
    \multicolumn{2}{c}{\textbf{RAA}} &
    \multicolumn{2}{c}{\textbf{RMC}} &
    
    \multicolumn{2}{c}{\textbf{Total}}
    \\

    % Midrules for the first header row
%    \cmidrule(lr){1-1}
    \cmidrule(lr){2-3}
    \cmidrule(lr){4-5}
    \cmidrule(lr){6-7}
    \cmidrule(lr){8-9}
    \cmidrule(lr){10-11}
    \cmidrule(lr){12-13}
    \cmidrule(lr){14-15}
    \cmidrule(lr){16-17}

    % Second Header Row: Column Titles
    \multicolumn{1}{l}{\rotatebox[origin=c]{0}{\textbf{Project}}} &
%    \multicolumn{1}{l}{\shortstack{\#Mut}} &
    \multicolumn{1}{l}{\shortstack{\#Mut}} &
    \multicolumn{1}{l}{\shortstack{MS(\%)}} &
    \multicolumn{1}{l}{\shortstack{\#Mut}} &
    \multicolumn{1}{l}{\shortstack{MS(\%)}} &
    \multicolumn{1}{l}{\shortstack{\#Mut}} &
    \multicolumn{1}{l}{\shortstack{MS(\%)}} &
    \multicolumn{1}{l}{\shortstack{\#Mut}} &
    \multicolumn{1}{l}{\shortstack{MS(\%)}} &
    \multicolumn{1}{l}{\shortstack{\#Mut}} &
    \multicolumn{1}{l}{\shortstack{MS(\%)}} &
    \multicolumn{1}{l}{\shortstack{\#Mut}} &
    \multicolumn{1}{l}{\shortstack{MS(\%)}} &
    \multicolumn{1}{l}{\shortstack{\#Mut}} &
    \multicolumn{1}{l}{\shortstack{MS(\%)}} &
    \multicolumn{1}{l}{\shortstack{\#Mut}} &
    \multicolumn{1}{l}{\shortstack{MS(\%)}}
    \\

    \midrule
    % Data Rows with Project Stats and Mutation Data

\newtag{1}{row:blinker}. \href{https://github.com/pallets-eco/blinker}{blinker} & 23 & 70 & 0 & NA & 1 & 0 & 0 & NA & 38 & 92 & 3 & 67 & 7 & 100 & 72 & 83\\
\newtag{2}{row:schedule}. \href{https://github.com/dbader/schedule}{schedule} & 2 & 50 & 0 & NA & 0 & NA & 0 & NA & 2 & 100 & 0 & NA & 2 & 100 & 6 & 83\\
\newtag{3}{row:pyjwt}. \href{https://github.com/jpadilla/pyjwt}{pyjwt} & 24 & 100 & 0 & NA & 0 & NA & 0 & NA & 38 & 100 & 5 & 100 & 18 & 100 & 85 & 100\\
\newtag{4}{row:pyquery}. \href{https://github.com/gawel/pyquery}{pyquery} & 93 & 85 & 5 & 80 & 2 & 100 & 6 & 67 & 62 & 95 & 12 & 100 & 111 & 75 & 291 & 84\\
\newtag{5}{row:funcy}. \href{https://github.com/Suor/funcy}{funcy} & 150 & 100 & 18 & 100 & 0 & NA & 1 & 100 & 67 & 100 & 17 & 100 & 5 & 100 & 258 & 100\\
\newtag{6}{row:wtforms}. \href{https://github.com/pallets-eco/wtforms}{wtforms} & 207 & 81 & 22 & 77 & 0 & NA & 1 & 100 & 303 & 97 & 31 & 100 & 85 & 80 & 649 & 89\\
\newtag{7}{row:pythonpatterns}. \href{https://github.com/faif/python-patterns}{python\_patterns} & 9 & 78 & 0 & NA & 4 & 75 & 1 & 100 & 36 & 92 & 13 & 100 & 18 & 94 & 81 & 91\\
\newtag{8}{row:marshmallow}. \href{https://github.com/marshmallow-code/marshmallow}{marshmallow} & 21 & 81 & 2 & 100 & 0 & NA & 7 & 43 & 13 & 77 & 0 & NA & 6 & 100 & 49 & 78\\
\newtag{9}{row:structlog}. \href{https://github.com/hynek/structlog}{structlog} & 141 & 61 & 13 & 100 & 4 & 75 & 6 & 67 & 204 & 89 & 41 & 98 & 22 & 100 & 431 & 81\\
\newtag{10}{row:Flask}. \href{https://github.com/pallets/flask}{flask} & 324 & 62 & 5 & 60 & 8 & 0 & 0 & NA & 290 & 89 & 46 & 89 & 145 & 87 & 818 & 77\\
\newtag{11}{row:graphene}. \href{https://github.com/graphql-python/graphene}{graphene} & 7 & 100 & 9 & 100 & 0 & NA & 0 & NA & 10 & 100 & 8 & 100 & 9 & 100 & 43 & 100\\
\newtag{12}{row:drf}. \href{https://github.com/encode/django-rest-framework}{django-rest-framework} & 4 & 100 & 4 & 100 & 2 & 100 & 0 & NA & 4 & 100 & 7 & 100 & 9 & 100 & 30 & 100\\
\newtag{13}{row:praw}. \href{https://github.com/praw-dev/praw}{praw} & 370 & 32 & 47 & 40 & 0 & NA & 1 & 100 & 656 & 96 & 6 & 100 & 130 & 75 & 1,210 & 72\\

     \midrule
     Mean & 106 & 77 & 10 & 84 & 2 & 58 & 2 & 82 & 133 & 94 & 15 & 96 & 44 & 93 & 309 & 88 \\

    \bottomrule
    \end{tabular}
    }
    
%\vspace{-8pt}
\end{table*}

Our goal was to show that \tool generates complementary mutants that exhibit behavioural diversity, and reveal faults overlooked by general-purpose mutation operators.
To do so, we compared mutation scores from both tools and analyzed cross-kill rates, test overlap, and test-mutant correlation. These analyses address the following sub-questions:

\begin{itemize}
    \item \textbf{RQ2.1}: How do the mutation scores of \tool and \comparedTool compare across the benchmark projects?

    \item \textbf{RQ2.2}: How many mutants are unique to each tool based on test outcomes and subsumption metrics established in prior literature? 

    \item \textbf{RQ2.3}: How strongly are the mutants from each tool coupled with the same test cases? What does this reveal about their fault model complementarity?
\end{itemize}

\paragraph{Experimental Object} Cosmic Ray is a lightweight, parallel mutation testing tool for Python. It supports a broad set of mutation operators, including statement deletions, constant replacements, and control-flow changes, and is widely adopted in the Python ecosystem~\cite{DBLP:conf/issta/FortunatoC022, DBLP:conf/dsa/DialloCWL24, 10.1145/3701625.3701659}.

\paragraph{Mutation Scores} To address \textbf{RQ2.1}, we executed \comparedTool on 11 of the 13 benchmarks and computed mutation scores using their original test suites. We excluded ``python\_patterns'' and ``praw'' due to compatibility issues with \comparedTool.
Similar to RQ1, we removed Cosmic Ray’s syntactically- and runtime-invalid mutants before comparison to ensure fairness.

\paragraph{Mutant Subsumption}
To address \textbf{RQ2.2}, we evaluated subsumption relationships between mutants generated by \tool and \comparedTool  by inferring a dynamic mutant subsumption graph~\cite{10.1109/ICSTW.2014.20}. Given the availability of test suites in mutation testing, this concept provides a practical alternative to undecidable true subsumption~\cite{10.1109/ICSTW.2014.20} and over-approximating static subsumption~\cite{DBLP:conf/icst/KurtzAO15}, and is formally defined as follows.

\begin{definition}[\emph{\textbf{Dynamic Subsumption Graph}}]\label{def:subsume::graph}
Let \(M\) be a finite set of mutants and \(T\) a finite set of tests. A mutant \(m \in M\) is said to \emph{dynamically subsume} another mutant \(m' \in M\) if there exists a test \(t \in T\) that kills \(m\) and every test in \(T\) that kills \(m\) also kills \(m'\). 
The \emph{dynamic subsumption graph} \(G\) is a directed graph whose nodes are mutants in \(M\) and whose edges represent this relation.
\end{definition}

We compute this graph over the combined mutants from both tools.
For each mutant in \tool, we compare its kill set against all \comparedTool mutants to detect subsumption.
Mutants not subsumed by any counterpart are marked \textit{unique}; and others \textit{subsumed}. 
We repeat the process symmetrically for both tools to quantify behavioural overlap and uniqueness.

We additionally computed the \textit{cross-kill rate} to quantify behavioural redundancy:

\[
{\small
\text{Cross-Kill Rate}(A,B)=
}
{\scriptsize
\frac{|\text{Mutants}_{A,B}^{\text{killed}}|}
     {|\text{Mutants}_{A}^{\text{killed}}| + |\text{Mutants}_{B}^{\text{killed}}| - |\text{Mutants}_{A,B}^{\text{killed}}|}
}
\]

\begin{comment}

\begin{equation*}
    \resizebox{\columnwidth}{!}{
    $\text{Cross-Kill Rate}(A,B) =
    \frac{|\text{Mutants}_{A,B}^{\text{killed}}|}
         {|\text{Mutants}_{A}^{\text{killed}}| + |\text{Mutants}_{B}^{\text{killed}}| - |\text{Mutants}_{A,B}^{\text{killed}}|}$
    }
\end{equation*}

\end{comment}

Here, $|\text{Mutants}_{A}^{\text{killed}}|$ and $|\text{Mutants}_{B}^{\text{killed}}|$ represent the number of killed mutants from tools $A$ and $B$, respectively, while $|\text{Mutants}_{A,B}^{\text{killed}}|$ is the number of mutants that are shared across both tools and are killed by tests. We used our previous assessment of subsumed mutants to identify the shared mutants.
A low cross-kill rate suggests behavioural diversity, while a high rate indicates redundancy.

\paragraph{Test Overlap and Coupling}
For \textbf{RQ2.3}, we studied how strongly mutants from each tool were coupled to the same test cases.
We computed the \textit{test overlap ratio}, defined as the proportion of tests that killed mutants from both tools:

\[
{\small
\text{Test Overlap Ratio}(A,B)=
}
{\scriptsize
\frac{|\text{Tests}_{A,B}^{\text{killed}}|}
     {|\text{Tests}_{A}^{\text{killed}}|
     + |\text{Tests}_{B}^{\text{killed}}|
     - |\text{Tests}_{A,B}^{\text{killed}}|}
}
\]

%Here, 
%$|\text{Tests}_{A,B}^{\text{killed}}|$ is the number of test cases that kill mutants from both tools $A$ and $B$, while
%$|\text{Tests}_{A}^{\text{killed}}|$ and $|\text{Tests}_{B}^{\text{killed}}|$ are the number of tests that kill mutants from each tool, respectively. A lower test overlap ratio suggests greater differences in the fault models of tools $A$ and $B$.

We also built a contingency table of test-mutant interactions and computed Pearson and Spearman correlations~\cite{de2016comparing, hauke2011comparison}, along with \textit{Cramér’s V}~\cite{cramer1999mathematical}, to assess the association between test behaviour across tools.
These analyses reveal whether \tool's mutants trigger distinct testing behaviour and therefore add new fault-detection capabilities.

\subsubsection{Experiment 3}

To address \textbf{RQ3}, we evaluated the performance overhead of \tool by measuring the time taken during each major phase of execution: (i) mutation candidate identification, (ii) mutation injection and test execution, and (iii) post-processing and report generation. For each benchmark, we averaged the timings across three independent runs to ensure consistency.
All experiments were conducted on a machine equipped with an Apple M2 processor (8-core @ 3.49GHz), running macOS 14.5, and 16GB of RAM.

\subsection{Results and Discussion}
\label{sec:results}

\subsubsection{RQ1}

\autoref{tab:mutations} presents the results of RQ1.
Each pair of columns reports the number of mutants and corresponding mutation score per operator.
For example, columns two and three indicate the count and mutation score for \RFAFull (\RFA) mutants, respectively.
The last two columns summarize the total mutants and overall score per project.
\tool generated an average of 309 mutants per project, with an average mutation score of 88\% (72--100\%).

The \textbf{lowest mutation scores} were observed for \RECFull (\REC), \DECFull (\DEC), \RFAFull (\RFA), and \RCFFull (\RCF),
with scores as low as 58--84\%, indicating test suites struggled to detect these faults.
To understand these deficiencies, we analyzed surviving mutants.
Dynamic typing in Python often hinders the detectability of faults.
This was especially evident for \RCF mutants: removing type conversion functions led to data type changes
that frequently went undetected ($\sim$16\% survived), unlike in statically-typed languages where such issues would be caught at compile time.
Another weakness stemmed from Python's support for default values and variadic arguments,
making it harder for tests to detect missing arguments.
$\sim$23\% of the \RFA mutants survived, despite argument removal.
Complex  conditions were also inadequately tested.
$\sim$18\% of the \DEC mutants remained undetected.
This may in part stem from Python’s coverage tools not reporting branch coverage by default, unlike tooling for Java or C{++}.

We then examined operators with the \textbf{highest mutation scores}, including \CUAFull (\CUA), \DAAFull (\DAA), and \RMCFull (\RMC), indicating more effective test coverage.
Despite significantly changing runtime behaviour, some mutants still survived due to Python's permissiveness.
In \RMC, for example, removing a method call sometimes left a valid expression: Python treated the object as \textit{truthy} even without the call.

We observed \textbf{recurring patterns} among surviving mutants.
Pythonic idioms, though concise and expressive, often impaired testability.
This aligns with prior findings on the challenges of testing idiomatic Python code~\cite{DeIdiom}.

Consider the following simplified example from \textit{Funcy}, a functional toolkit for Python (Table~\ref{tab:benchmarks}, row~\ref{row:funcy}). %\footnote{Simplified for clarity.}
The \code{signature_repr()} function generates a string representation of a function’s signature, primarily for debugging.
It extracts the name (line~\ref{line:funcy_name}) and arguments (line~\ref{line:funcy_args}). If the function accepts \code{**kwargs}, it processes them via a generator expression (line~\ref{line:funcy_mutant}).

\begin{lstlisting}[language=python, escapechar=|]
def signature_repr(call): |\label{line:funcy_def}|
    name = get_func_name(call) |\label{line:funcy_name}|
    args_repr = (repr(arg) for arg in call.args) |\label{line:funcy_args}|
    kwargs_repr = ('%s=%s' % (key, smart_repr(value, repr_len)) for key, value in call._kwargs.items()) |\label{line:funcy_mutant}|
    return '%s(%s)' % (name, ', '.join(chain(args_repr, kwargs_repr))) |\label{line:funcy_return}|
\end{lstlisting}

\tool applied \REC to mutate this line by removing the \code{value} element from \code{<key, value>} pairs.
Despite 96\% test coverage, the mutant survived. %, indicating the output format was not validated by tests.
This illustrates that even well-covered code can lack behavioural assertions, especially when idiomatic Python is used.

We observed poor testing of exception handling and debugging logic.
Tests often neglected edge cases, focusing on main execution paths, echoing prior work on robustness gaps~\cite{Dalton20}.
Several projects exhibited inflated global coverage due to module imports, though mutants in such regions survived.

\paragraph{Equivalent Mutants}
\tool generated \AvgEquivPerPrj equivalent mutants on average per project, ranging from \AvgEquivRange.
\citet{Guerino2024} reported an average of 5.7\% (std-dev=7.6\%) equivalents for Cosmic Ray.
Our heuristics pruned \AvgHeuristicDetected of candidates as likely equivalent.
%After removing equivalents, \tool’s mutation score increased from \AvgMSPerPrj to \NewMSTool.
%
The few equivalent mutants mainly stemmed from \REC and \RCF.
In \REC, some removed elements were unused, making mutations semantically neutral.
In \RCF, mutants often remained equivalent due to implicit type coercions (e.g., \code{str(x)} replaced by \code{x} before a \code{print()}).

%\annotate{meta comment 8: add invalid mutant statistics (R1.C3)}%
\paragraph{Invalid mutants}
We analyzed syntactic- and runtime-invalid mutants as indicators of operator robustness.
As shown in the one-before-last column of \autoref{tab:comparison_rq2}, most projects produced few or none, while three (\emph{Flask}, \emph{Marshmallow}, and \emph{Schedule}) accounted for most invalids in \tool.
These projects rely heavily on dynamic composition and late binding, often causing unresolved attributes or container references after mutation.
Syntactic invalids were rare in \tool, consistent with Python’s simple AST transformations; most arose at runtime.
Overall, invalids formed only a small fraction of mutants.
Table~\ref{tab:invalids} summarizes \tool's invalids per operator, with most stemming from \textit{ChUsedAttr} and \textit{RemElCont}, which target highly dynamic features.

 \begin{table}[t]
 \centering
 \small
 \caption{\tool's invalid mutants per mutation operator.}
 \label{tab:invalids}

 \begin{tabular}{lrrr}
 \toprule
 \textbf{Operator} & \textbf{Invalids} & \textbf{\% per MO} & \textbf{\% of Total} \\
 \midrule
 ChUsedAttr & 313 & 15.6 & 7.05 \\
 RemAttrAcc & 39 & 15.2 & 0.88 \\
 RemElCont & 78 & 66.1 & 1.76 \\
 RemExpCond & 15 & 55.6 & 0.34 \\
 RemFuncArg & 46 & 3.5 & 1.04 \\
 RemMetCall & 77 & 12.7 & 1.73 \\
 RemConvFunc & 9 & 7.1 & 0.20 \\
 \bottomrule
 \end{tabular}
 
% \vspace{-10pt}
 \end{table}

\infobox{
\tool revealed faults missed by existing Python test suites, including in high-coverage projects.
It introduced realistic, Python-specific faults that eluded tests, exposing key gaps in coverage and test oracles.
}

\subsubsection{RQ2}

\autoref{tab:comparison_rq2} presents the results for RQ2. On average, \tool generated  248 mutants, while \comparedTool produced 2,537, after validation.
The last column of~\autoref{tab:comparison_rq2} reports the number of invalid mutants removed from \comparedTool, which were also concentrated in a small subset of projects and operators, but remained a small fraction overall.
		
\paragraph{RQ2.1. Mutation Scores}
Columns 4–5 in~\autoref{tab:comparison_rq2} show the mutation scores. \tool achieved an average mutation score of 89\% (range: 77--100\%), while \comparedTool averaged 67\% (range: 0--100\%).

\begin{table*}[!t]
\caption{Mutant information for PyTation (PyTa) and Cosmic Ray (CoRa): 
valid mutants (\#Mut), mutation scores (MS), percentage of unique mutants, 
cross-kill rate, test-overlap ratio, and invalid mutants.}
\label{tab:comparison_rq2}
\centering
\scriptsize
\resizebox{\textwidth}{!}{
\begin{tabular}{l rr rr rr r r rr}
\toprule
\textbf{Project} 
& \multicolumn{2}{c}{\textbf{\#Mut}} 
& \multicolumn{2}{c}{\textbf{MS(\%)} }
& \multicolumn{2}{c}{\textbf{Unique Mut(\%)}}
& \textbf{Cross-Kill(\%)} 
& \textbf{Test Overlap(\%)} 
& \multicolumn{2}{c}{\textbf{\#Invalid}}
\\
\cmidrule(lr){2-3}
\cmidrule(lr){4-5}
\cmidrule(lr){6-7}
\cmidrule(lr){10-11}

& PyTa & CoRa 
& PyTa & CoRa 
& PyTa & CoRa 
&  
&  
& PyTa & CoRa 
\\
\midrule

\newtag{1}{row:blinker}. \href{https://github.com/pallets-eco/blinker}{blinker}  & 72 & 310 & 83 & 0 & 100 & 100 & 0.00 & 0.00 & 0 & 0 \\

\newtag{2}{row:schedule}. \href{https://github.com/dbader/schedule}{schedule} & 6 & 600 & 83 & 89 & 50 & 85 & 0.38 & 3.75 & 178 & 0 \\

\newtag{3}{row:pyjwt}. \href{https://github.com/jpadilla/pyjwt}{pyjwt} & 85 & 2112 & 100 & 35 & 44 & 88 & 4.95 & 10.20 & 0 & 133 \\

\newtag{4}{row:pyquery}. \href{https://github.com/gawel/pyquery}{pyquery} & 291 & 1271 & 84 & 68 & 56 & 56 & 13.30 & 23.40 & 2 & 21 \\

\newtag{5}{row:funcy}. \href{https://github.com/Suor/funcy}{funcy} & 258 & 2034 & 100 & 62 & 53 & 75 & 9.69 & 39.57 & 4 & 43 \\

\newtag{6}{row:wtforms}. \href{https://github.com/pallets-eco/wtforms}{wtforms} & 649 & 1195 & 89 & 88 & 26 & 46 & 8.82 & 20.41 & 9 & 68 \\

\newtag{7}{row:marshmallow}. \href{https://github.com/marshmallow-code/marshmallow}{marshmallow} & 49 & 3341 & 78 & 0 & 100 & 100 & 0.00 & 0.00 & 98 & 0 \\

\newtag{8}{row:structlog}. \href{https://github.com/hynek/structlog}{structlog} & 431 & 378 & 81 & 100 & 97 & 100 & 0.14 & 0.42 & 23 & 0 \\

\newtag{9}{row:Flask}. \href{https://github.com/pallets/flask}{flask} & 818 & 4034 & 77 & 100 & 100 & 0 & 1.19 & 0.77 & 2 & 8 \\

\newtag{10}{row:graphene}. \href{https://github.com/graphql-python/graphene}{graphene} & 43 & 6694 & 100 & 100 & 35 & 98 & 0.22 & 3.12 & 0 & 281 \\

\newtag{11}{row:drf}. \href{https://github.com/encode/django-rest-framework}{django-rest-framework} & 30 & 5941 & 100 & 100 & 100 & 100 & 0.00 & 0.00 & 1 & 5 \\

\midrule
mean & 248 & 2537 & 89 & 67 & 69 & 77 & 3.52 & 9.24 & 29 & 51 \\
\bottomrule

\end{tabular}
}
\end{table*}

These results reflect the complementary strengths of both tools.
Both tools uncovered alive mutants in global scopes and edge-case logic, reinforcing the RQ1 observation that high coverage does not guarantee behavioural adequacy.
\tool's lower but still substantial mutation score, alongside its high rate of unique mutants, indicates that its Python-specific operators provide added value by uncovering test suite deficiencies not exposed by general-purpose mutations.

\paragraph{RQ2.2. Mutant Subsumption}
On average, 69\% of mutants generated by \tool are not dynamically-subsumed by any mutants from \comparedTool, while \comparedTool produces 77\% unique mutants. This suggests that the two tools generate mutants that trigger largely distinct runtime behaviours.
This underscores \tool’s ability to capture fault-revealing behaviours missed by traditional mutation operators.

The average cross-kill rate between \tool and \comparedTool mutants was 3.52\%, indicating that few \tool mutants were killed by the same tests as \comparedTool's. 
These findings support the premise that our anti-pattern-based mutation operators target distinct fault models.
While general-purpose operators are effective, \tool’s Python-specific mutants complement them by covering fault types underrepresented in existing approaches.
    		
\paragraph{RQ2.3. Mutant Coupling and Test Overlap}
The test overlap ratio (third-to-last column of \autoref{tab:comparison_rq2}) between \tool's unique mutants and those of \comparedTool averaged 9.24\%, suggesting that test cases killing \comparedTool's mutants rarely killed those from \tool.
We also constructed a test-mutant contingency table 
(tests × mutants, killed/not killed) and computed Pearson and Spearman correlations, along with Cramér’s~V on 2×2 tool–outcome tables.
No statistically significant association ($p>0.05$) was found, supporting that \tool’s mutants elicit distinct test behaviours compared to \comparedTool.
%\deletedtext{This lack of dependency further supports that the two tools target different fault spaces.}

Because cross-kill is sensitive to differing mutant counts, we interpret it together with test overlap and correlation results.
Test overlap offers a coarse but complementary view of test–mutant coupling, indicating how often tests reveal shared faults.
We adapt conditionally-overlapped mutants~\cite{ma2016mutation}, simplifying output comparison to a binary killed/alive status.

\paragraph{Example}

We examined a mutation  (lines~\ref{line:structlog_class}--\ref{line:structlog_mutation}) from \textit{structlog} (Table~\ref{tab:benchmarks}, row \ref{row:structlog}), a logging framework.
\comparedTool mutated the constructor of the \texttt{MaybeTimeStamper} class by replacing \texttt{|} with \texttt{<<} in the type hint, a change that was caught by the test suite.
\tool, by contrast, applied the \RFA operator and removed the optional \texttt{utc} argument, causing the constructor to default to UTC timestamps, instead of the local time. This mutant survived, revealing a missed behavioural check not covered by existing tests, despite the same scope being mutated by both tools.

\begin{lstlisting}[language=python, escapechar=@]
class MaybeTimeStamper: @\label{line:structlog_class}@
    def __init__(self, fmt: str | None = None, utc: bool = True): @\label{line:structlog_init}@
        self.stamper = TimeStamper(fmt=fmt, @\hl{utc=utc}@) @\label{line:structlog_mutation}@
\end{lstlisting}

We also observed that \tool introduced mutants in smaller functions or return expressions that \comparedTool did not mutate. For instance, \tool's argument-deletion operators applied to return statements uncovered missed faults in concise utility functions.
Overall, \tool introduced mutants in code scopes untouched by \comparedTool in \textasciitilde17\% of cases, further demonstrating its complementary coverage.
       
        \infobox{
        Our findings demonstrate that \tool enhances Python mutation testing by producing behaviourally distinct mutants that reveal faults overlooked by general-purpose tools like \comparedTool. The low  cross-kill and test overlap rates highlight that \tool exercises different test behaviours, confirming its value in exposing diverse fault scenarios and improving test suite assessment.
        }

\subsubsection{RQ3}
    % Results
    The results of \textbf{RQ3} showed that the average execution time of \tool was \AVGTime and the median
    was \MEDTime on each benchmark with exercising the tests on the mutants being the most time-consuming phase.
    On average, \tool spent \AVGIdentificationTime identifying the mutants (\AVGStaticTime for static analysis
        and \AVGDynamicTime for dynamic analysis), the mutation process allocated \AVGMutationTime, and
    the post-processing time was \AVGPostProcessTime, which is negligible.
    Worst case time was \MaxTime for \MaxTimePrj project and the best case time was \MinTime for \MinTimePrj project.
Normalized by mutant count, \tool’s mean runtime (25.2 s/mutant) was comparable to Cosmic Ray’s (24.8 s/mutant), also showing minimal overhead from dynamic analysis ($\sim$3.5 m/project). Both tools support parallel execution, further mitigating runtime cost.

\subsubsection{Threats to Validity}
\label{sec:threats}

A key internal threat is examiner bias during manual classification of equivalent mutants.
This labour-intensive process relies on interpretation of program behaviour and may introduce subjectivity, a limitation common in equivalent-mutant studies~\cite{Mirshokraie13}.
To mitigate bias, two researchers independently reviewed the results.
An external threat concerns the representativeness of benchmarks and test suites.
We selected 13 open-source projects spanning diverse sizes, architectures, and domains, comparable to or larger than prior mutation studies~\cite{Delgado17MuCPP, Mirshokraie13, Derezinska11, PIT, 10962485}.
Each project was evaluated using its developer-written test suite, a common practice for assessing real-world rather than synthetic tests.
Our dynamic analysis introduces test-suite dependence in mutant placement (not killability), a trade-off shared with other coverage-guided mutation tools.
To support reproducibility, we publicly released the source code and benchmark suite used in this study.

\section{Related Work}

Traditional coverage criteria help assess test adequacy~\cite{Zhu97} and correlate with fault detection when controlled for test size~\cite{Namin09}, but not strongly with test effectiveness~\cite{Inozemtseva14}. Mutation testing, which evaluates a test suite’s ability to detect injected faults~\cite{10.5555/909408, Offutt2001}, has proven effective for improving testing~\cite{Jia11, Just14, Papadakis19, Petrovic21, Sanchez24}, though scalability and cost remain challenges~\cite{Papadakis10, pizzoleto20, 10.1145/2483760.2483782, delgado2017assessment}.
\textbf{Equivalent mutants} are a core obstacle~\cite{Offutt2001}, addressed using coverage heuristics~\cite{schuler13}, state-based classification~\cite{Papadakis14Eq}, constraint solving~\cite{Offut97}, symbolic execution~\cite{Baer20}, near-correct input generation~\cite{Schwander21}, and AST-based models~\cite{Just17}.
\textbf{Language-specific mutation tools} have advanced testing in Java~\cite{PIT, Javalanche, Offutt2004, Kintis17}, C++\cite{Delgado17MuCPP}, C\#\cite{Derezinska11}, and WS-BPEL~\cite{Estero08, 5463642}, often by focusing on object-oriented features. For dynamic languages, Bottaci~\cite{Bottaci10} proposed runtime-type-guided mutation, extended to Smalltalk~\cite{Gligoric11}. JavaScript-specific strategies include domain operators~\cite{Mirshokraie13} and NLP-based context mutation~\cite{Ritcher22}.

\textbf{Python mutation and analysis.} Derezińska and Hałas~\cite{Derezinska14} proposed static operators excluding type-dependent cases. Open-source tools like Mutmut~\cite{Mutmut}, Cosmic Ray~\cite{Cosmicray}, and MutPy~\cite{Mutpy} apply general-purpose operators and use only coverage data. Python dynamic analysis tools include DynaPyt~\cite{DynaPyt} (instrumentation), PyCG~\cite{PyCG} (static call graphs), PyTER~\cite{PyTER} (patch synthesis), and Mypy~\cite{Mypy} (type inference). Fault studies like PySStuBs~\cite{PySStuBs} and BugsInPy~\cite{Bugsinpy} catalogue common bugs.
None of these tools target Python-specific anti-patterns using both static and dynamic signals. Unlike our approach, they either apply generic mutations or lack runtime-guided filtering to reduce equivalent mutants. Our technique fills this gap by combining language-aware mutation design with dynamic analysis to guide and refine mutation testing.

%\section{Related Work}

\section{Concluding Remarks}\label{sec:conclusion}

Python’s dynamic nature complicates mutation testing and limits purely static operators.
We presented a Python mutation testing approach based on anti-pattern-driven operators and hybrid static--dynamic analysis.
Across 13 real-world projects, \tool exposed test inadequacies missed by a state-of-the-art tool, generating realistic and diverse mutants with low redundancy.
Our results highlight that Python’s dynamic features frequently introduce untested behaviours overlooked by general operators.
Future work includes test-suite augmentation and integrating mutation data with fault localization to support debugging~\cite{fault_localization}.

\begin{acks}
This work was supported in part by a Natural Sciences and Engineering Research Council of Canada (NSERC) Discovery Grant.
We thank the ICSE reviewers for their careful reading and constructive feedback.
We thank A.~de~Vaal for assistance with early experimentation on this work.
We also acknowledge G.~Neisi~Minaei and N.~Modares~Gorji for developing an initial prototype codebase.
\end{acks}

\balance
\bibliographystyle{ACM-Reference-Format}
\bibliography{src/refs}

\end{document}